\documentclass[acmsmall]{acmart}
\pdfoutput=1
\AtBeginDocument{\providecommand\BibTeX{{\normalfont B\kern-0.5em{\scshape i\kern-0.25em b}\kern-0.8em\TeX}}}

\newtheorem{problem}{Problem}

\usepackage{multirow}
\usepackage{subcaption}
\usepackage{wrapfig}
\usepackage{todonotes}
\usepackage{numprint}\npthousandsep{,}\npthousandthpartsep{}\npdecimalsign{.}

\theoremstyle{definition}
\newtheorem{defn}{Definition}[section]

\renewcommand\footnotetextcopyrightpermission[1]{}
\begin{document}

	\title{On the Aggression Diffusion Modeling and Minimization in Twitter}
	
	\author{Marinos Poiitis}
	\email{mpoiitis@csd.auth.gr}
	\orcid{0000-0002-9810-2907}
	\affiliation{%
		\institution{Aristotle University of Thessaloniki}
		\city{Thessaloniki}
		\country{Greece}
	}
	
	\author{Athena Vakali}
	\email{avakali@csd.auth.gr}
	\orcid{1234-5678-9012}
	\affiliation{%
		\institution{Aristotle University of Thessaloniki}
		\city{Thessaloniki}
		\country{Greece}
	}
	
	\author{Nicolas Kourtellis}
	\email{nicolas.kourtellis@telefonica.com}
	\orcid{1234-5678-9012}
	\affiliation{%
		\institution{Telefonica Research}
		\city{Barcelona}
		\country{Spain}
	}
	
	\renewcommand{\shortauthors}{Poiitis, et al.}
	
	\begin{abstract}
		Aggression in online social networks has been studied mostly from the perspective of machine learning which detects such behavior in a static context. However, the way \textit{aggression diffuses} in the network has received little attention as it embeds modeling challenges.
		In fact, modeling how aggression propagates from one user to another, is an important research topic since it can enable effective aggression monitoring, especially in media platforms which up to now apply simplistic user blocking techniques. 
		
		In this paper, we address aggression propagation modeling and minimization in Twitter, since it is a popular microblogging platform at which aggression had several onsets. We propose various methods building on two well-known diffusion models, \textit{Independent Cascade} ($IC$) and \textit{Linear Threshold} ($LT$), to study the aggression evolution in the social network. We experimentally investigate how well each method can model aggression propagation using real Twitter data, while varying parameters, such as seed users selection, graph edge weighting, users' activation timing, etc. It is found that the best performing strategies are the ones to select seed users with a degree-based approach, weigh user edges based on their social circles' overlaps, and activate users according to their aggression levels. We further employ the best performing models to predict which ordinary real users could become aggressive (and vice versa) in the future, and achieve up to $AUC$=0.89 in this prediction task. Finally, we investigate aggression minimization by launching competitive cascades to ``inform'' and ``heal'' aggressors. We show that $IC$ and $LT$ models can be used in aggression minimization, providing less intrusive alternatives to the blocking techniques currently employed by Twitter.
		
	\end{abstract}

	
	
	\maketitle

	\section{Introduction}
	\label{sec:Introduction}
	
	Online social media offer unprecedented communication opportunities, but also come with unfortunate  malicious behaviors. Cyberbullying, racism, hate speech and discrimination are some of the online aggressive behaviors manifesting in such platforms and often have devastating consequences for individual users and the society as a whole. Aggression can be explicit through inappropriate posting such as negative feelings and embarrassing photos, or implicit when it unconsciously hurts online users (e.g., through negative gossip spreading). Overall, online social media users are often left exposed and vulnerable to potential aggression threats. 
	
	Inter-disciplinary studies have focused on cyber aggression from the perspective of social psychology and computational sciences. Social learning and bonding~\cite{allen2017general, smith2008cyberbullying}, as well as the theory of planned behavior~\cite{lee2002holistic} provide the basis of its theoretical formulation. Furthermore, machine learning has been used to detect such behavior in online platforms (e.g.~\cite{chatzakou2017mean, davidson2017automated, founta2018large, waseem2016hateful}). Even with all this body of earlier work, online aggression has not been uniformly defined~\cite{corcoran2015cyberbullying}. Hence, online aggression is formulated under varying approaches, depending on the severity of the aggressive behavior, the type of platform and social interactions it facilitates, the power of the aggressor over the victim etc.
	
	Interestingly, aggressive behavior has been found to be tightly related to the social circle of the individual expressing it~\cite{huang2014cyber}, while its origins are located in the aggressive peer influence and the effect of the initial aggressors on their network~\cite{runions2015online}, suggesting a potential \emph{cascading spreading trend} from the source of the behavior to its connections. Additionally, the above findings have been expanded by analyzing the underlying network structure and the pair-wise interactions, further enhancing the intuition that \emph{online aggression propagates akin to a diffusion process}. Therefore, the aggression's overall effect can be strengthened as it propagates through the network~\cite{henneberger2017effect}. Surprisingly, aggression \textit{propagation} in the cyber-space has gathered little attention, primarily due to the complexity and dynamic nature of the problem. Specifically, the overall diffusion process is influenced by factors explicit to the specifics of each social network, i.e, how users form connections with each other, as well as network-agnostic ones such as the effect of user anonymity on the aggression manifestation~\cite{zimmerman2016online}. This hardship is evident from the lack of automated processes in the popular media platforms, such as Twitter, to mitigate aggression's negative effects, which are restricted to blocking or reporting abusive users and removing inappropriate content~\cite{basak2019online}.
	
	The present work focuses on the \textit{online aggression propagation problem} and studies the complexities of modeling and minimizing aggression propagation in online social networks. It particularly focuses on Twitter, which apart from being one of the largest social media platforms, it openly provides user data that have been already analyzed by several studies within the context of aggression.
	It considers the well known diffusion models of {\textit Independent Cascading (IC)} and {\textit Linear Threshold (LT)}~\cite{kempe2003maximizing} as they provide the building blocks for studying the diffusion process at its two fundamental types of interactions: user-user ($IC$) and user-neighborhood ($LT$). To this end, we formulate aggression-aware information diffusion through appropriate parameters and enable a thorough study of the aggression dynamics. Then, we show how $IC$ and $LT$ models can be used in aggression minimization, providing less intrusive alternatives to the techniques currently in use by Twitter.
	
	\begin{figure}[htbp]
		\centering
		\includegraphics[width=\textwidth]{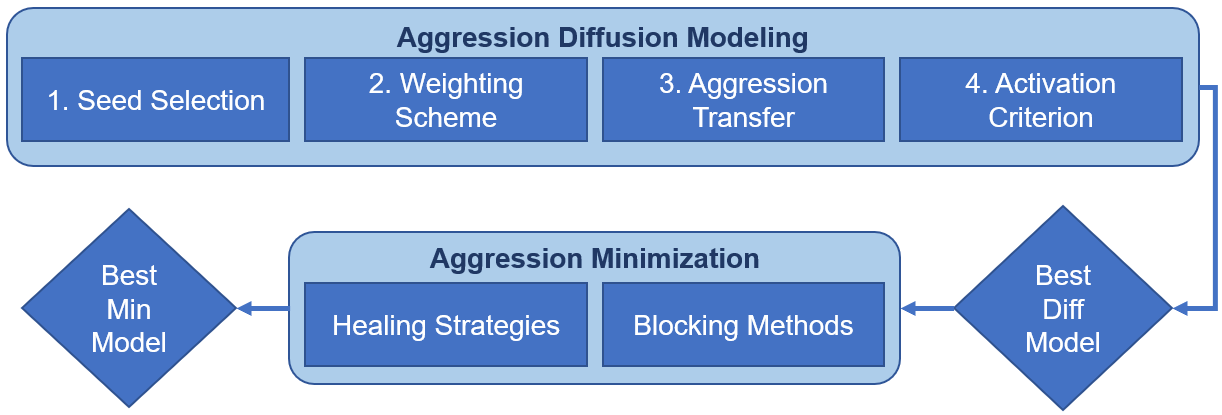}
		\caption{The overall process of the current work. Initially, aggression modeling is studied by setting and experimenting on its various sub-components, including selection of seed nodes, weighting scheme of edges, transfer of aggression and activation of nodes. The best models can then be used to minimize the overall network's aggression with further configurations by applying healing strategies. Finally, the best methods are compared to the currently used blocking mechanisms.}
		\label{fig:overview}
	\end{figure}
	
	The overall process is outlined in Figure~\ref{fig:overview}. Initially, aggression diffusion modeling is examined, where its sub-components - seed selection strategy, weighting scheme, activation criterion/threshold strategy and aggression transferring - are defined and explored. The process leads to two models, an $IC$ and an $LT$-based, with the best expressive performance. Next, aggression minimization step exploits these two models to define and examine the method of competitive cascades and its dynamics (i.e., healing strategy) as an alternative to the blocking methods, which are currently used by Twitter, to reduce the overall aggression level of the social network.
	
	To address the above considerations regarding online aggression diffusion and minimization, the main contributions of this work are:
	\begin{enumerate}
		\item \textbf{C1: aggression diffusion theoretical foundation,}\label{contribution1} by exploring $IC$ and $LT$ as the basic propagation models. Upon them, the theoretical notions of user-user interaction, initial user selection strategies and user propagation are introduced and adjusted accordingly. The same models are also utilized in aggression minimization methods, where various healing approaches are formulated (Section~\ref{sec:methodology}).
		\item \textbf{C2: aggression modeling and minimization experimentation,}\label{contribution2} for both $IC$ and $LT$, through similarity performance tests, statistical validation as well as aggression reduction specifically for the minimization process. Real data experimentation, extensive simulations, a modeling case study and comparison to the blocking minimization methods currently in use by Twitter and similar online social platforms support our findings (Section~\ref{sec:results}).
		\item \textbf{C3: Exploration on results and implications}\label{contribution3}, by validating the experimental results for both modeling and minimization. Specifically, neighborhood similarity is shown to be the most appropriate criterion according to which user relationships are formed, while central users are the best to initiate the diffusion process. Finally, the adjusted competitive cascades are shown to outperform current Twitter banning methods achieving a reduction of $\sim$50\% for $IC$ and $\sim$15\% for $LT$ (Section~\ref{sec:results}).
		\item \textbf{C4: Simulation framework release} \label{contribution4} for reproducibility purposes and further experimentation or extensions\footnote{https://anonymous.4open.science/r/95063f31-d3aa-4171-80d7-e0efb890476d/}.
	\end{enumerate}
	
	The rest of this paper is organized as follows. In Section~\ref{relatedWork} the related works are reviewed. Section~\ref{sec:methodology} provides the theoretical foundation of both aggression modeling and minimization. The experimental evaluation is presented in Section~\ref{sec:results} and Section~\ref{sec:future} provides a discussion of this work's results and further improvements.
	
	\section{Related Work}
	\label{relatedWork}
	
	Information diffusion has been originally studied by probabilistic models~\cite{domingos2001mining, richardson2002mining}, which were advanced by considering a  discrete optimization problem formulation driven by the well-known Independent Cascade ($IC$) and Linear Threshold ($LT$) diffusion models~\cite{kempe2003maximizing}.
	These models capture not only the neighborhood influence, but also the cumulative influence of a user's social circle where online aggression phenomena can occur~\cite{henneberger2017effect}.
	
	\textit{Opinion Dynamics ($OD$)} is another family of information diffusion models which studies the effects and conditions of information propagation over the network.
	Under $OD$, neighborhood influence is crucial for the diffusion process and it is manifested through pairwise user interactions~\cite{fernandez2014voter}. Another set of models are the well known \textit{Susceptible, Infected, Recovered (SIR)} models from the field of epidemics. They are a generalization of $IC$, categorizing users into the above three states and studying the transitions from one state to another, under specific thresholds~\cite{trpevski2010model}. Both OD and SIR are useful in understanding the dynamics of diffusion process and determining proper aggression propagation models. However, $OD$ methods serve as one of many approaches to online aggression modeling, while as discussed below in Section~\ref{agg_modeling}, there is evidence suggesting the eligibility of models capturing mass information dissemination, such as $IC$ and $LT$. Last, a more generic investigation of the various $SIR$ models is voluminous and left as future work.
	
	\subsection{Influence Maximization}
	
	\textit{Influence Maximization} proposes a real-world alternative to the above models. Specifically, under some limitations in resources - and precisely in the number of seed nodes that the propagation process could be initiated from~\cite{tong2017adaptive} - it aims to maximize the expected influence in the network~\cite{borodin2010threshold}.
	
	As a result, \textit{Influence Maximization} aims at optimizing the seed selection process by minimizing the selection cost~\cite{chen2009approximability, long2011minimizing}, or even sustaining the spread of the cascade above a specified threshold~\cite{long2011minimizing}. For this purpose, the greedy approach is shown to outperform methods based on network and node properties~\cite{kempe2003maximizing}, however many optimization methods have been proposed as the simple greedy version is computationally infeasible. In particular \textit{Cost-Effective Lazy Forward selection} ($CELF$) leverages the mathematical set property of submodularity, achieving an 700\% improvement in execution time~\cite{leskovec2007cost}, while \textit{Single Discount (SD)} and \textit{Degree Discount (DD)} heuristics outperform $CELF$ in terms of execution time by more than six orders of magnitude while maintaining a high influence spread~\cite{chen2009efficient}. In particular, $SD$ finds the most central nodes based on an iterative selection and removal process, while $DD$ is fine-tuned towards $IC$ model and exploits its user activation probability. In the opposite direction than the degree heuristics, Morone et al.~\cite{morone2015influence} propose the usage of low degree and high betweenness nodes as seeds, based on the intuition that degree by itself is not sufficient to capture a node's importance. Instead, these special bridge nodes are shown to be highly efficient as they are crucial for the information flow despite their loose connectivity.
	
	Besides $IC$ and $LT$ models, \textit{Simulated Annealing} optimizes the spread function of the diffusion process both in terms of time and magnitude through a local search algorithm~\cite{jiang2011simulated}, while heat diffusion is utilized to simulate information diffusion~\cite{ma2008mining}. These natural approaches introduce the decaying effect of time on a user's influence and they have largely inspired the \textit{Decaying Aggression Transfer} healing mechanism proposed in our work.
	
	\subsection{Competitive Cascades}
	
	In both information diffusion and influence maximization, the presence of a parallel, sometimes negative cascade has been mentioned. Under this scenario, each cascade selects a seed set to initiate its propagation and depending on the assumed hypotheses, they either compete to activate a node or they spread regardless of the competitor and measure the impact at the end of both processes. These are known as \textit{Competitive Cascades} and help us formulate the first part of our proposed aggression minimization process.
	
	\textit{Competitive Influence Maximization:}
	The most usual task of competitive cascades is to find a positive seed set that minimizes the spread of the negative cascade~\cite{wu2017scalable, he2012influence}. This problem is called \textit{Influence Blocking Maximization (IBM)} and has been addressed with both $IC$ and $LT$. However, these methods retrace to the discussed intractable greedy algorithm, thus they either propose tailor-made problem variations of IBM~\cite{wu2017scalable}, or they introduce model specific structures, such as the \textit{Local Directed Acyclic Graph}, applicable only on $LT$~\cite{he2012influence}. Furthermore, \textit{preference} for $LT$ and \textit{authorities} for $IC$ have been introduced to address simultaneous influence of both cascades~\cite{zhang2015limiting, budak2011limiting, wu2017scalable}. These notions are useful in this work's competitive aggression minimization process.
	
	\subsection{Immunization}
	
	Social media nowadays tend to address abusive behavior restriction by blocking abusive users or posts. This preference is grounded on either the simpler nature of those algorithms, or due to their more straightforward application. Hence, apart from the competitive cascades, the latest \textit{immunization} or \textit{blocking} algorithms are also reviewed in this work and compared to discover the most efficient approach in terms of aggression reduction.
	
	State-of-the-art methods exploit the largest eigenvalue $\lambda$ of the network adjacency matrix \textbf{A} which has been shown to be the only network parameter that determines whether the diffusion would become an epidemic or not~\cite{prakash2012threshold}. Specifically, \textit{NetShield} algorithm introduces a ranking score called \textit{Shield score} to detect the most important abusive users~\cite{tong2010vulnerability}. Furthermore, NetMelt transfers the problem to edge deletion, as node deletion is more radical and intrusive~\cite{tong2012gelling}. However, such methods operate offline, i.e., they capture the network state before the propagation process starts, in contrast to competitive cascades that we propose.
	
	\subsection{Aggression Modeling and Minimization}
	\label{agg_modeling}
	
	Research on aggression has been focused on its detection rather than its modeling. Specifically, Twitter account features have been used to detect and categorize phenomena such as hate speech~\cite{waseem2016hateful}, while incremental crowdsourcing techniques have been proposed to annotate abusive tweets~\cite{founta2018large}. Additionally, graph features such as peer pressure and cumulative influence have also been highlighted~\cite{squicciarini2015identification}. Furthermore, a multi-class classifier has been used for the classification among various types of abusive behavior like racism or homophobia~\cite{davidson2017automated}, while Random Forests have been exploited to detect disturbing phenomena such as cyberbullying~\cite{chatzakou2017mean}.
	
	In the context of online aggression detection, there have been some works investigating the dynamics of the online aggressive behavior, its persistence and spreading. More concretely, a study on Twitter data, addresses the local ego-networks to formulate the user neighborhood and finds that the online social influence is of significant importance to the detection of cyberbullies~\cite{huang2014cyber}. Expanding this work to a macroscopic view, the underlying graph structure and pair-wise user interactions are investigated leading to the importance of peer pressure on the spread of cyberbullying over the social network~\cite{squicciarini2015identification}. Last, to theoretically solidify these practical approaches, Runions et al. suggest that in online social networks the reason of the aggressive behavior's adoption is the influence of the person who originated the content~\cite{runions2015online}. Despite all this volume of works on cyberbullying and more generally on online aggression detection, the clarification of the aggression dynamics and properties seem more like a side-effect instead of a concrete goal.
	
	According to the authors' knowledge, there is only a single approach modeling online aggression diffusion which exploits $OD$ models~\cite{terizi2020angry} in Twitter. While $OD$ can simulate the neighborhood influence through user-user interactions, prior research in Psychology suggests the need for a simultaneous and mass scale diffusion process to capture online aggression. In particular, it is found that social bonds with computer abusers significantly motivate people to commit computer abuses~\cite{lee2002holistic}, while these victim users can be influenced by their social circle to act aggressively not only through pairwise interactions with members of this circle but also through a mass and synchronous interaction with the whole neighborhood~\cite{henneberger2017effect}. Last, through a more applied perspective, it is proposed to employ aggregated models for global cyberbullying dynamics to capture this cascading mass effect~\cite{squicciarini2015identification}. 
	Also, to the best of our knowledge,\textit{ we are the first to address online aggression minimization. In particular, two different minimization methods are explored here and applied on Twitter using: (a) competitive cascades, (b) node/edge blocking which are compared with respect to aggression reduction efficiency.}
	
	\section{Methodology}
	\label{sec:methodology}
	
	In this section, we address and formalize the process pipeline and notions presented in Figure~\ref{fig:overview}. In particular, we provide the theoretical foundations for user interactions, seed selection, aggression propagation and healing mechanisms, and connect them to the diffusion and minimization problems, as per \textit{Contribution~\ref{contribution1}}. 
	
	\subsection{Aggression Modeling}
	\label{sec:agg_modeling}	
	A social network is represented by a directed graph $G=(V,E)$, where $V$ is a set of n nodes or users, and $E$ is a set of m edges, i.e., relationships between users. Each node is associated with an aggression score denoted by $A_i$ $\forall v_i \in V$. The aim is to find whether the $IC$ and $LT$ are able to model aggression diffusion and what are the important parameters that enable them to do so.
	
	\subsubsection{Seed Selection}
	\label{sec:seedselect}
	Under aggression diffusion modeling, the seed selection process is the first crucial component to describe. \textit{Seed} nodes are the ones the diffusion process starts. The ultimate result of the propagation model is tightly related to these initial seed nodes. The necessity for selecting seed nodes is due to the diffusion model being restricted by a given budget $k$, meaning that we can only afford a specific amount of initial stimulation of $k$ seed nodes. As a result, a sophisticated seed selection strategy could enable the diffusion process and lead to the increase of the number of activated nodes.
	
	Even though the model as a whole relies mostly on the continuous node aggression scores, it is beneficial to define what an aggressive node is distinguishing them from the normal ones, either for future work or to address a binary version of the problem. The definition of the aggressive node given below will be proved useful in the formulation of the various seed strategies:
	
	\begin{defn}
		Given the set of aggression scores \textbf{A}, and a node $u$ with aggression score $A_u$, $u$ is called \textit{aggressive} iff $A_u \geq A_{Pi\%}$, where $A_{Pi\%}$ is the aggression score of the i-th percentile node of the ranked distribution of \textbf{A}. 
	\end{defn}
	
	With all the necessary concepts defined and explained, the proposed seed strategies are presented below:
	
	\begin{itemize}
		\item \textbf{All Aggressive:} Set all aggressive nodes as seed nodes.
		\item \textbf{Top Aggressive:} From the set of aggressive nodes, select the k nodes with the highest aggression score as seed nodes, where k is a user-defined parameter. If $k > |aggressive$ $nodes|$, use \textit{All Aggressive} strategy. 
		\item \textbf{Single Discount (SD):} Following prior work~\cite{chen2009efficient}, the process iteratively finds the node with the highest degree and places it in the seed set. Whenever a node $u$ is added to the seed set, all its neighbors' edges ending to $u$ are discarded, thus discounting the degree of $u$'s neighbors by 1 in the subsequent iteration. This is a basic heuristic applicable to all cascade models.
		\item \textbf{Degree Discount (DD):} Also presented in~\cite{chen2009efficient}, and in addition to the degree-based heuristic of $SD$, this approach exploits the node activation probabilities of the $IC$ model, to further discount nodes that are most likely to get affected by neighbors which are already contained in the seed set.
		\item \textbf{Betweenness-Degree (BD):} Sort nodes based on increasing degree and select the ones with betweenness centrality higher than the median value. This approach draws insights from previous work on influence maximization, where these important and simultaneously loosely connected nodes (low degree) play a crucial role in the information spreading process (high betweenness)~\cite{morone2015influence}.
		\item \textbf{Random:} Choose users as seeds randomly.
	\end{itemize}
	
	\subsubsection{Weighting Scheme}
	
	Next step is to define the various weighting schemes we apply on the network edges. Each scheme captures real world properties that can impact the aggression diffusion in a different way, as they can affect the probability of propagating aggression. In the following, we assume that the user is embedded in directional edges:
	
	\begin{itemize}
		\item \textbf{Jaccard overlap (Jaccard):} In social media, the friends and followers of a user can heavily influence the user's own beliefs. If two connected users appear to have similar or even identical social circles, this could point to very similar beliefs or behaviors. Therefore, given two connected nodes $u$ and $v$, and their corresponding sets of neighbors $N_u$ and $N_v$, the \textit{Jaccard overlap} of their edge is defined as weight $w_{uv} = \frac{N_u \cap N_v}{N_u \cup N_v} \in [0,1]$.
		
		\item \textbf{Power score (Power):} Given a node $u$, the \textit{Power score} $P_u$ is defined as the ratio of the in-degree (incoming edges) over out-degree (outgoing edges), $P_u = \frac{inDegree_u}{outDegree_u} \in [0,1]$, as the scores are ultimately normalized. The higher the \textit{Power score} of node $u$, the more dominant the influence it receives from its in-neighbors in comparison to the influence it applies on its out-neighbors. However, for this measure to be employed as an edge weight, it has to be considered in a pairwise fashion. Therefore, given an edge from $u$ to $v$, we define $P_{uv} = \frac{P_v}{P_u} \in [0,1]$ to capture $u$'s power on $v$. Qualitatively, $P_{uv}$ increases when $u$ becomes more influential or $v$ more prone to its neighbors' influence.
		
		\item \textbf{Weighted overlap (Weighted):} Given two nodes $u$ and $v$, their \textit{Jaccard overlap} $w_{uv}$ and their \textit{Power score} $P_{uv}$, the \textit{Weighted overlap} is defined as $Pw_{uv} = P_{uv} * w_{uv} \in [0,1]$. This metric combines the previous two weights.
	\end{itemize}
	
	\begin{figure}[t]
		\centering
		\begin{subfigure}[b]{0.32\textwidth}
			\centering
			\includegraphics[width=\textwidth]{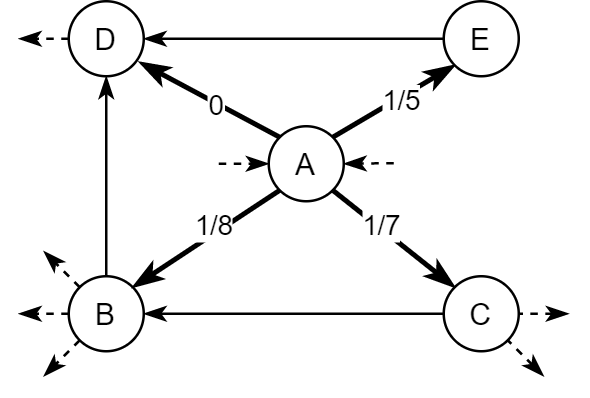}
			\caption{\textit{Jaccard}. (neighbors = successors)}
			\label{fig:ws_jaccard}
		\end{subfigure}
		\hfill
		\centering
		\begin{subfigure}[b]{0.32\textwidth}
			\centering
			\includegraphics[width=\textwidth]{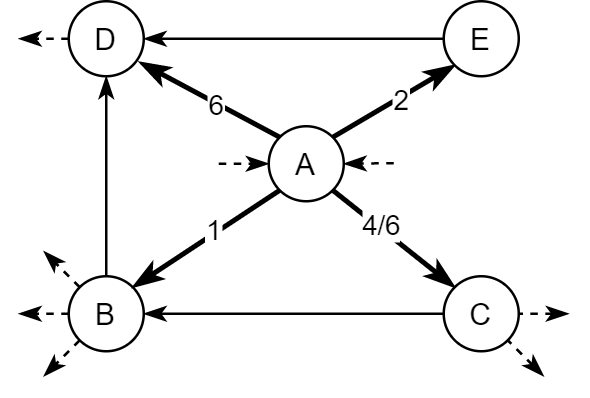}
			\caption{\textit{Power}}
			\label{fig:ws_power}
		\end{subfigure}
		\hfill
		\centering
		\begin{subfigure}[b]{0.32\textwidth}
			\centering
			\includegraphics[width=\textwidth]{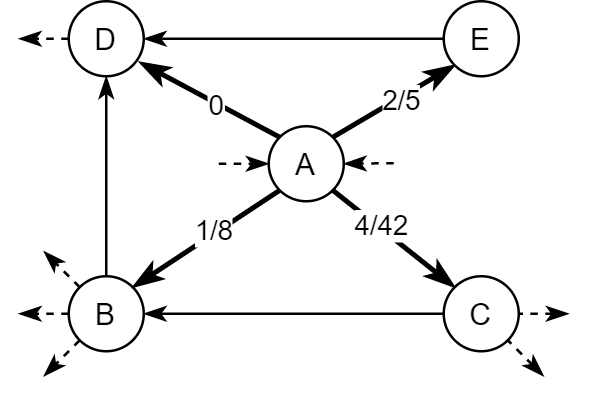}
			\caption{\textit{Weighted}}
			\label{fig:ws_weighted}
		\end{subfigure}
		\caption{Weighting scheme scenarios focusing on node A. Bold denotes the edges under examination and dashed edges refer to in/out-edges from/to the rest graph. Normalized scores are presented in parentheses.}
		\label{fig:ws_overall}
	\end{figure}
	
	\noindent
	Each one of the introduced weighting schemes is presented through a toy scenario in Figure~\ref{fig:ws_overall}. 
	Here, it is noted that the unnormalized weights are presented for the sake of clarity, while during the experimental process the normalized versions were employed (shown in parentheses).
	In Figure~\ref{fig:ws_jaccard}, \textit{Jaccard} is calculated for the bold edges using the neighbors of the source A and target nodes.
	For example, $w_{AB} = 1/8$ as A and B share one common neighbor from a total of seven. Notably, considering that the graph is directed, only the successor nodes are included in one's neighborhood.
	Proceeding, in Figure~\ref{fig:ws_power}, \textit{Power} scores are calculated using both in- and out-neighbors.
	Therefore, $P_{AB} = 1$, as both A and B have two in- and four out-neighbors.
	Last, \textit{Weighted} is depicted in Figure~\ref{fig:ws_weighted}, where $Pw_{AB} = P_{AB} * w_{AB} = 1/8 $ as it depends both on \textit{Jaccard} and \textit{Power}. Next, and using the above notations and metrics, we describe the $IC$ and $LT$ diffusion models, given a constant seed budget $k$.
	
	\subsubsection{Independent Cascade (IC)}
	
	At time step $t=0$ the process starts with an initial seed set $S$ of active nodes with $|S| \leq k$ and proceeds in discrete time steps according to the following stochastic rule. 
	At each time step $t$, a set of active nodes $Active^t$ are present in the network.
	Each active node $u \in Active^t$, that was activated in time step $t$, has a single chance to activate each one of its inactive neighbors, $v$, with probability $p$. This probability is defined with respect to the very same notions that were introduced in weighting schemes above. That is, node $v$ is activated by node $u$ with: $p \leq w_{uv}$, $p \leq P_{uv}$ or $p \leq Pw_{uv}$ when the activation criterion is \textit{Jaccard}, \textit{Power} or \textit{Weighted}, respectively.
	
	Under the aggressive $IC$ model, activation means that $u$ transfers its aggression score to user $v$, that is $A_v = A_u$. However, there is a case where multiple nodes $u \in N_v$ try to activate the same node $v$ simultaneously, and succeed. Then, the decision of whose aggression score to transfer follows one of the below strategies:
	
	\begin{itemize}
		\item \textbf{Random:} The activation order is arbitrary and the aggression score of a randomly selected node $u \in N_v$ from the successful ones (i.e., $N_v \cap Active^t$), is transferred to the node being activated $v$.
		\item \textbf{Top:} The most aggressive node among the successful ones transfers its aggression score to $v$.
		\item \textbf{Cumulative:} The cumulative aggression score of all nodes is transferred according to their contribution, a metric capturing peer pressure, also verified in~\cite{squicciarini2015identification}. For example, if $S=\{u_1, u_2, u_3\}$ the set of nodes that succeeded, then $A_v = \sum_{i \in S}w_i * A_i$ with $w_i = \frac{A_i}{\sum_{j \in S} A_j}$ and $A_i$ their respective aggression scores.
	\end{itemize}
	
	If node $u$ succeeds, then node $v$ gets activated in step $t+1$; but regardless of $u$'s success, it cannot make any further attempts to activate $v$ in a following step. The $IC$ process terminates when no more activations are possible.
	
	\begin{figure}[htbp]
		\centering
		\includegraphics[width=0.7\textwidth]{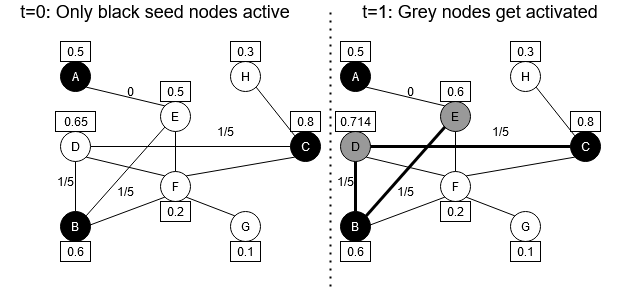}
		\caption{A single step of the aggression-based IC diffusion process. Node labels represent aggression scores, while edge labels are \textit{Jaccard} weights capturing the two participating nodes' neighborhood similarity. In this scenario, the \textit{cumulative} activation criterion is used, meaning that when two nodes activate a third node, they both contribute to its aggression score change, while the seed nodes are selected randomly. Bold lines denote the activators in this specific scenario.}	
		\label{fig:ic_diffusion}
	\end{figure}
	
	To clarify the aggression-related $IC$ diffusion process, Figure~\ref{fig:ic_diffusion} presents a single step of a given scenario. Specifically, in time step $t=0$, only the seed (black) nodes A, B and C are active, which have been picked randomly, i.e. according to \textit{Random} seed strategy. The edge weights have been defined based on \textit{Jaccard} weighting scheme, capturing the similarity between the neighborhoods of the two nodes participating in the corresponding edge, and only a sample of them is presented for clarity purposes. To give an example, the weight of the edge between nodes C and D is the number of common neighbors divided by the number of common and uncommon ones, i.e $\frac{1}{5}$. These very same weights act as the activation probability as well, meaning that an already active node could possibly activate the specific inactive neighbor with a probability equal to their edge's weight. On time step $t=1$, suppose that two new (gray) nodes - D and E - get activated. D is activated by two activators simultaneously (B and C), while E is activated by a single one (i.e., B), a process denoted by the bold edges. Here, it should be clarified that since the process is probabilistic, different nodes could have been activated by different combinations of activators in a different scenario. Continuing, in the case of E, the aggression score of B who activates E, is transferred as is to E, changing it from $0.5$ to $0.6$. In the case of D though, the \textit{Cumulative} activation criterion is used to break the tie as it is not clear which node among B and C should transfer its aggression. Therefore D adopts both of its activators' aggression scores, according to the definition of \textit{Cumulative} criterion above, meaning that D's aggression score becomes $\frac{0.8}{1.4} * 0.8 + \frac{0.6}{1.4} * 0.6 = 0.714$.
	
	\subsubsection{Linear Threshold (LT)}
	
	In the original $LT$ model, each node $v$ is assigned a threshold $\theta_v \in [0,1]$. The process starts with an initially activated seed set $S$ with $|S| \leq k$, similar to the $IC$ process. In step $t$, node $v$ is influenced by its already active neighbors $u \in N_v$ and gets activated if the ratio of its active neighbors exceeds the pre-specified and arbitrary threshold $\theta_v$.
	However, in the aggression-based $LT$, this activation strategy is an over simplification. In particular, instead of simply considering the fraction of active neighbors, aggression-based $LT$ utilizes the underlying weighting scheme to weight the influence of active neighbors. If this aggregated weighted influence exceeds the node's threshold, the node gets activated. 
	Thus, the original $LT$ model can be seen as a special case of the proposed model, applied on an unweighted graph. 
	Also, instead of assigning arbitrary thresholds to nodes, the thresholds here are related to a physical meaning, either node aggression or power score. Specifically, the alternative activation criteria at time step $t$ are: 
	
	\begin{itemize}
		\item \textbf{Aggression:} When aggression scores are used as node thresholds, that is:
		\[\theta_v = A_v \in [0, 1] \;,\;\; \forall v \in V\]
		\item \textbf{Power:} When power scores are used as node thresholds, that is:
		\[\theta_v = P_v \in [0, 1] \;,\;\; \forall v \in V\]
	\end{itemize}
	Then, node $v$ gets activated if:
	\[\theta_v \leq \sum_{u \in N_v\cap Active^t}w_{e_{vu}},\]
	where $w_{e_{vu}}$ is the weight of edge $e_{vu}$.
	Finally, under the aggressive $LT$ model, and in contrast to the $IC$ case, activation has a single interpretation: the active neighbors of node $v$ transfer their average aggression score to $v$, that is $A_v = \frac{\sum_{u \in \bar{N}_v}A_u}{|\bar{N}_v|} \in [0,1]$, where $\bar{N}_v$ is the set of active neighbors of $v$. In this context, the process unfolds until there are no nodes left to become active according to the thresholds.
	
	\begin{figure}[htbp]
		\centering
		\includegraphics[width=0.7\textwidth]{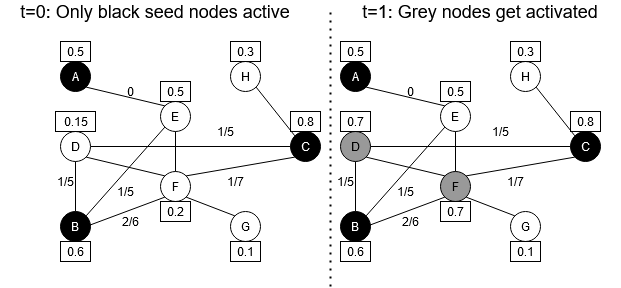}
		\caption{A single step of the aggression-based LT diffusion process. Node labels represent aggression scores while edge labels are \textit{Jaccard} weights, capturing the two participating nodes' neighborhood similarity. In this scenario, \textit{Aggression} threshold strategy is used, meaning that nodes transfer their aggression explicitly based on their tendency towards it instead of some other attribute of theirs, while the seed nodes are selected randomly.}
		\label{fig:lt_diffusion}
	\end{figure}
	
	Figure~\ref{fig:lt_diffusion} presents a single step of a scenario for the aggression-based $LT$ diffusion model. Similar to the case of the $IC$ scenario, the seed (black) nodes A, B and C initiating the diffusion process have been randomly selected. Also, \textit{Jaccard} has been applied to weigh the graph, by capturing the neighborhood similarity of two connected nodes, a notion that is very useful considering that $LT$ exploits a node's neighborhood influence by definition. Additionally, \textit{Aggression} threshold strategy is enabled, forcing inactive nodes to adopt the aggression of their activators based on their own aggressiveness instead of some other metric or feature they may have at their disposal. $LT$ model differentiates from the $IC$ on two crucial mechanisms. First, no tie braking rule is necessary, as all the active neighbors contribute equally in the aggression transferring process. Second, the process is not stochastic, meaning that given the specific scenario only a single outcome is possible, as there are no probabilities involved, just deterministic thresholds. This is why there are no bold edges here, in contrast to the ones appearing in the $IC$ scenario.
	For completeness of the example, we examine time step $t=1$ of the figure, when the two seed nodes B and C activate their (grey) neighbors D and F.
	These are to be activated because their thresholds are low enough. The activation is executed by transferring both of their aggression scores, expressed as the average of the two values. Specifically, D's threshold (which is its aggression score $A_D=0.15$) is lower than its B and C neighbors' influence, calculated as the average of the participating edge weights $\frac{1/5 + 1/5}{2} = 0.2$. Therefore, this leads to D's aggression score to change from $0.15$ to $\frac{0.6 + 0.8}{2} = 0.7$. In a similar manner, F is updated from $0.2$ to $0.7$. 
	
	\subsection{Aggression Minimization}
	
	Given the two models ($IC$ and $LT$), we address the problem of aggression minimization in Twitter by means of two different approaches, called \textit{competitive} and \textit{blocking} aggression minimization, respectively.
	
	\subsubsection{Competitive Aggression Minimization (CAM)}
	
	Under this minimization method, there are two competing diffusion processes, a negative - also called aggressive - and a positive or educational cascade.
	The goal of the latter is to minimize the spread of the negative one. For the negative cascade and hence the aggression diffusion, we use the best model discovered in aggression modeling (as explained earlier), while for the positive one, we use the corresponding cascading model and test the alternative configurations.\\
	
	\begin{problem}[CAM]
		
		Given a directed network $G=(V,E)$, with $V$ and $E$ denoting the node and edge sets, respectively, an integer budget $k$ and two competing diffusion processes, the aggressive and educational, with the latter having a set of parameters $\theta (\cdot)$, the $CAM$ problem is to find a positive seed set $S$ with $|S| \leq k$ and $\theta^\ast(\cdot)$, s.t. $\theta^\ast(\cdot) = \arg\min_\theta A(G)$ w.r.t. $S$, where $A(G) = \sum_{v \in V} A_v$ is the overall aggression score in the network.
	\end{problem} 
	
	We now define the rules that describe the educational cascade. Under $IC$, the activation probabilities follow the same notions of the negative cascade with respect to weighting schemes: \textit{Jaccard overlap}, \textit{Power score} or \textit{Weighted overlap}. However, under $LT$, the activation criterion as well as node thresholds should depend on \textit{Power} only. \textit{Aggression} would be an inappropriate threshold as the educational cascade intents to mitigate the effect of aggression on the network, instead of maximizing the overall aggression score.
	
	Using a predefined negative seed set, and according to the best strategy of aggression modeling scenario, we allow the two processes to unfold simultaneously. If at time step $t$, both cascades reach the same node $v$, then $v$ gets positively activated, simulating the fact that most probably an educated or aware user would stop manifesting aggressive behavior.
	
	Moving to the meaning of activation from the perspective of the positive cascade - the healing effect - it is noted that the seed nodes of the positive cascade present a special reaction to the educational process. That is, regardless of the seed strategy that is followed to select them, they are considered to be the most sensitized as the educational piece of information initiates its propagation from them. Hence, the healing effect on them should be drastic and independent of the rest of the diffusion process. Specifically, the positive seed nodes' aggression score is immediately zeroed out providing that the educational information does affect them. To simulate this process, a positive seed is totally healed with a probability $p$, depending on its original aggression score. In the other case, the seed node neglects this information and sustains its aggression. The more aggressive the seed node, the more difficult for the healing to occur.
	More formally, the new aggression score of the positive seed nodes follows the below rule: 
	
	\[
	A'_s = \begin{cases}
		0 & \text{if } p \geq A_s\\
		A_s & otherwise
	\end{cases}
	\]
	where $A_s$ is the initial aggression score of node $s$, and $A'_s$ its score after the healing effect has occurred. Apart from seed nodes, we discern four possible cases:
	
	\begin{itemize}
		\item \textbf{Vaccination}: activating user $v$ results to $A'_v \rightarrow 0$, i.e., the user becomes normal immediately. This hypothesis is strict and serves as an upper bound to the minimization's performance as it captures the ideal case of a user becoming completely sensitized by the educational cascade without ever reverting to the aggressive behavior.
		\item \textbf{Aggression transfer}: activating user $v$ results to:
		\begin{itemize}
			\item \textbf{IC}: $A'_v \rightarrow A_u$, with user $u$ activating $v$
			\item \textbf{LT}: $A'_v \rightarrow \frac{\sum_{u \in N_v}A_u}{|N_u|}$
		\end{itemize}
		This is the case where aggression is transferred according to the underlying mechanics of the diffusion model, as described in Section~\ref{sec:agg_modeling}. Therefore, reducing the overall aggression of the network depends solely on the influence of the positive cascade's seed nodes, considering they are the only ones starting with a reduced aggression as described above.
		\item \textbf{Decaying transfer}: activating user $v$ is affected by a decaying factor $\lambda$ capturing the distance from the source of the information. Hence, $\lambda = \frac{1}{hops}$ and:
		\begin{itemize}
			\item \textbf{IC}: $A'_v \rightarrow \lambda*A_u$, with user $u$ activating $v$
			\item \textbf{LT}: $A'_v \rightarrow \frac{\sum_{u \in N_v}\lambda * A_u}{|N_u|}$
		\end{itemize}
		This mechanism simulates the fading effect of the positive cascade, meaning that users immediately exposed to it are more prone to change in comparison to those getting affected indirectly e.g. by a friend's re-posting activity. 
		\item \textbf{Hybrid}: using a combination of cases $1$ and $3$, this case aims to capture the different potential reactions of the various users when they become exposed to the positive cascade. Formally:
		\[
		A'_v = \begin{cases}
			\text{Vaccination} & \text{, if } p \geq A_v\\
			\text{Decaying transfer} & , otherwise
		\end{cases}
		\]
	\end{itemize}
	
	Finally, with respect to the seed strategy of the positive cascade, two alternatives are considered. On one hand, the seed strategy of the negative cascade can be used, but this scenario favors the negative cascade as the decision would be based on its own specifics. On the other hand, one of the other strategies proposed in aggression modeling section can be exploited, regardless of the seed strategy of the negative cascade.
	
	\subsubsection{Blocking Aggression Minimization (BAM)}
	
	In contrast to CAM, in \textit{Blocking Aggression Minimization} (BAM) there is only a single cascade, the aggressive one. The aim here is to target specific nodes or edges to immunize. 
	That means removing them to optimally suppress aggression diffusion over the network. In the case of social networks, node removal is equal to banning a user which is a drastic measure. This is why we consider the case of edge removal also. Formally the problem is defined below:
	
	\begin{problem}[BAM-N](node version)
		
		Given a directed network $G=(V,E)$, with $V$ and $E$ denoting the node and edge sets, respectively, and an integer budget $k$, the problem is to find a subset of nodes $S \subseteq V$ with $|S| = k$, s.t. $S = \arg\min_{s \in P(V)}A(G)$, where $A(G) = \sum_{v \in V} A_v $ is the overall aggression score in the network, and $P(V)$ is the Power Set of $V$.
	\end{problem} 
	
	\begin{problem}[BAM-E](edge version)
		
		Given a directed network $G=(V,E)$, with $V$ and $E$ denoting the node and edge sets, respectively, and an integer budget $k$, the problem is to find a subset of edges $S \subseteq E$ with $|S| = k$, s.t. $S = \arg\min_{s \in P(E)}A(G)$, where $A(G) = \sum_{v \in V} A_v $ is the overall aggression score in the network, and $P(E)$ is the Power Set of $E$.
	\end{problem}
	
	Based on the major finding of~\cite{prakash2012threshold} and~\cite{wang2003epidemic}, for a large family of diffusion processes, the only network parameter that determines whether this diffusion would become an epidemic or not is the largest eigenvalue $\lambda$ of the adjacency matrix \textbf{A}. For this reason, in our experiments next, we use \textit{NetShield}~\cite{tong2010vulnerability} to solve the \textit{BAM-N} problem, and \textit{NetMelt}~\cite{tong2012gelling} for \textit{BAM-E} problem. Motivated by the results of these two studies, coupled with the competitive minimization explained earlier, we also implement an aggression-related variation. In particular, instead of using the initial edge weights, we exploit the product of the aggression scores of source and destination nodes.
	That, is for pair $(u,v)$ in the adjacency matrix, the cell value now becomes $A_u * A_v$.
	However, it should be noted that $BAM$ is a problem that exploits offline methods, in contrast to $CAM$ which adjusts to the dynamics of the aggression diffusion. Concluding, the above problems are tested and validated in the experimental section below. 
	
	\section{Experimental Results}\label{sec:results}
	
	In this section, and as per our \textit{Contribution~\ref{contribution2}}, we present the results of the experimental process, first with respect to aggression modeling using $IC$ and $LT$ methods (Sec.~\ref{sec:aggression-modeling-experiments} and~\ref{sec:case-study}), and then aggression minimization using competing cascades or blocking (Sec.~\ref{sec:aggression-minimization-experiments}).
	
	\subsection{Experimental Setup}
	\label{experimental_setup}
	
	\textbf{Dataset:}
	Despite the limited but yet available datasets used in the online aggression context (i.e. MySpace and FormSpring~\cite{squicciarini2015identification}), to the best of our knowledge, currently there is no existing dataset providing time-related snapshots of the aggression information, a trait which is a critical part of our analysis. Thus, in this work, we utilize two different Twitter datasets, a \textbf{small labeled (SL)}~\cite{chatzakou2017mean} and a \textbf{large unlabeled (LU)}~\cite{mcauley2012learning}. By using SL, a Random Forest classification algorithm~\cite{terizi2020angry} is trained that is then used to label LU, a process which is briefly described below.
	
	SL is constructed by using 401 users as seeds for crawling their ego-network, to gather their friends and followers. Specifically, two crawls were made, one in 2016 and one in 2019. For the first time snapshot in 2016, the 401 initial users were originally annotated as \textit{normal, spammers, bullying} or \textit{aggressive}, but to better match this work's setup, \textit{spammers} are removed and \textit{bullying} users are merged with \textit{aggressive} into a single \textit{aggressive} class. However, for the crawled users no state is available.
	Thus, it was conservatively supposed that they were all normal.
	For the second snapshot in 2019, the active and deleted users are labeled as \textit{normal}, while the newfound suspended accounts as \textit{aggressive}, following the findings of prior work~\cite{chatzakou2019detecting}.
	Overall, the SL dataset consists of \numprint{1171218} users labeled as \textit{normal} or \textit{aggressive} and \numprint{1787543} edges between them, from which \numprint{477113} edges were dropped since their endpoints were not retrievable.
	This dataset consists: (a) the ground truth for training the Random Forest classifier of ~\cite{terizi2020angry} to annotate the LU graph, as presented below, and (b) the case study presented in Section~\ref{sec:case-study}.
	
	LU is comprised of \numprint{81306} users and \numprint{1768149} directed edges between them~\cite{mcauley2012learning}. To reduce noise in the converging process, we use the strongly connected component of this network, which consists of \numprint{68413} nodes and \numprint{1685163} directed edges. The edge weights are decided according to the proposed weighting schemes described in Section~\ref{sec:methodology}. In order to acquire labels to bootstrap our propagation algorithms, following similar methodology as in~\cite{terizi2020angry}, we apply the Random Forest prediction algorithm proposed in~\cite{chatzakou2017mean}, which provides a probability for a user to be \textit{aggressive} or not, based on their characteristics in the network and their activity. We tune this algorithm to use only the available network-related features and a total of 10 generated trees and train it on the SL ground truth dataset, leading to an overall $93.2\%$ in terms of accuracy, precision and recall. Last, the classifier is employed to compute the initial user aggression scores of all users in LU, which is equal to the probability of a user being aggressive. Ultimately, \numprint{5594} users (or $\sim$8\% of the network) were given a non-zero aggression score.
	
	\begin{table}[t]
		\begin{minipage}[b]{.25\textwidth}
			\centering
			\caption{\label{table:validation_vector}Validation vector transition features}
			\resizebox{\textwidth}{!}{\begin{tabular}[t]{c | c }
					\hline
					$t_0$ & $t_i$\\
					\hline \hline
					n & \{n $||$ a$\}_i$\\
					a & \{n $||$ a$\}_i$\\
					N-N & \{N-N $||$ N-A $||$ A-N $||$ A-A$\}_i$\\
					N-A & \{N-N $||$ N-A $||$ A-N $||$ A-A$\}_i$\\
					A-N & \{N-N $||$ N-A $||$ A-N $||$ A-A$\}_i$\\
					A-A & \{N-N $||$ N-A $||$ A-N $||$ A-A$\}_i$\\
					
			\end{tabular}}
		\end{minipage}
		\hfill
		\begin{minipage}[b]{.69\textwidth}
			\centering
			\caption{Cosine similarity for $LT$ models, with different Weighting Schemes, Seed Strategies and Threshold Strategies. * means that the reported cosine similarity is insensitive to the specific parameter's values, given the rest of the configuration parameters. The best value is highlighted in bold.}
			\label{table:lt_results}
			\resizebox{\textwidth}{!}{\begin{tabular}[t]{c | c | c | c}
					\hline
					\textbf{Weighting Scheme} & \textbf{Seed Strategy} & \textbf{Threshold Strategy} & \textbf{Cosine Similarity}\\
					\hline \hline
					
					\textit{Jaccard} & \textit{Random | $BD$} & * & 0.690\\
					\textit{Jaccard} & \textit{All Aggressive} & * & 0.688\\
					\textit{Jaccard} & $SD$ & \textit{Aggression} | \textit{Power} & \textbf{0.691} | 0.690\\
					\textit{Power} & * & * & 0.689\\
					\textit{Weighted} & * & * & 0.689\\
					
			\end{tabular}}
		\end{minipage}
	\end{table}
	
	\noindent
	\textbf{Metrics used to measure aggression change:}
	Similar to~\cite{terizi2020angry}, we measure the state of aggression of users, and how it changes through simulated time using a vector of $26$ metrics. The 6 core metrics enlisted below capture the network state with respect to users and edges and their label at time $t_i$:
	
	\begin{itemize}
		\item \textbf{n}: portion of normal users in the network
		\item \textbf{a}: portion of aggressive users in the network
		\item \textbf{N-N}: portion of edges that both users $i$ and $j$ are normal
		\item \textbf{N-A}: portion of edges that user $i$ is normal \& $j$ aggressive
		\item \textbf{A-N}: portion of edges that user $i$ is aggressive \& $j$ normal
		\item \textbf{A-A}: portion of edges that $i$ and $j$ are aggressive users 
	\end{itemize}
	
	By combining these core metrics, Table~\ref{table:validation_vector} presents 20 additional metrics capturing the transitions through simulated time between $t_i$ and initial state $t_0$. This metrics vector applied on the SL dataset for which there are labels available, forms the ground truth vector against which all of the IC and LT configurations are tested during the experimental phase. The actual values of the ground truth vector were provided by~\cite{terizi2020angry}.
	
	\noindent
	\textbf{Comparing simulation and ground truth data:}
	The above set of metrics is computed for all simulated models and all their time steps and compared with the ground truth vector using Cosine Similarity, i.e., \textit{Cosine(ground\_truth\_vector, simulation\_vector)}. In fact, we tried other similarity metrics to capture the agreement with ground truth, such as Pearson and Spearman correlations, but Cosine Similarity produced more stable results.
	
	The careful reader will notice that the state of users is binary (normal and aggressive). Hence, to compute the cosine of the two vectors at each simulated step, the users' state must be dichotomized. For this, we experimented with different thresholds ($T_A$) on the users' aggression scores ($T_A=\{0.1, 0.2, ..., 0.9\}$) and concluded that results with $T_A=0.4$ exhibit best overall similarity with ground truth data. Thus, all results presented next are based on $T_A=0.4$.
	
	
	\noindent
	\textbf{Statistical Tests:}
	Regarding $IC$ modeling, we run every experiment 10 times due to the inherent randomized nature of the activation process. Additionally, when \textit{Random} is used as the seed strategy, each experiment is executed 10 more times, leading to a total of 100 executions.
	To present these results, we use Cumulative Distribution Functions (CDF). To validate whether there is significant difference in parameter values, we employ \textit{One-Way ANOVA}, followed by a \textit{pairwise post-hoc Tukey's HSD} test, to spot the exact value of the significant parameter. We identify statistically significant differences at p-value $< 0.001$ for both tests.
	For the $LT$ modeling, we run each experiment only once, except for the case of \textit{Random} seed strategy, for which we execute each setup 10 times and acquire average performance. These results are presented in Table~\ref{table:lt_results}. The careful reader would observe that the various $LT$ configurations lead to slight changes in terms of cosine similarity. To validate the significance of the presented results, we once again rely to \textit{One-Way ANOVA} and \textit{Tukey's HSD} tests to identify which are the statistically different results. The insignificant ones are reported in Table~\ref{table:lt_results} with *.
	
	\noindent
	\textbf{Simulation Framework:}
	We designed and implemented a modular and extensible simulation framework to execute the modeling and minimization experiments.
	It is written in Python, it is open sourced, and allows fine-grained control of simulation parameters, to enable reproducible experiments, and future extensions (\textit{Contribution~\ref{contribution4}}).

	\subsection{Aggression Modeling}
	\label{sec:aggression-modeling-experiments}
	
	\begin{figure}[t]
		\centering
		\begin{subfigure}[b]{0.32\textwidth}
			\centering
			\includegraphics[width=\textwidth]{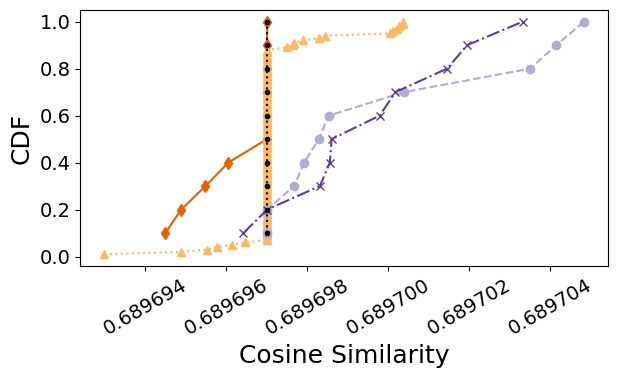}
			\caption{$WS$: Weighted; $AC$: Random}
			\label{fig:ic_seed_strategy_W}
		\end{subfigure}
		\hfill
		\centering
		\begin{subfigure}[b]{0.32\textwidth}
			\centering
			\includegraphics[width=\textwidth]{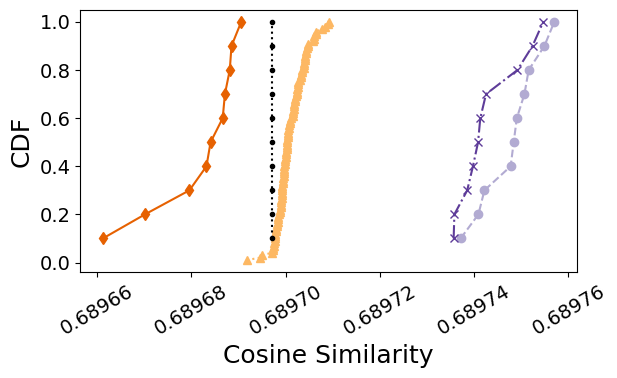}
			\caption{$WS$: Power; $AC$: Cumulative}
			\label{fig:ic_seed_strategy_P}
		\end{subfigure}
		\hfill
		\centering
		\begin{subfigure}[b]{0.32\textwidth}
			\centering
			\includegraphics[width=\textwidth]{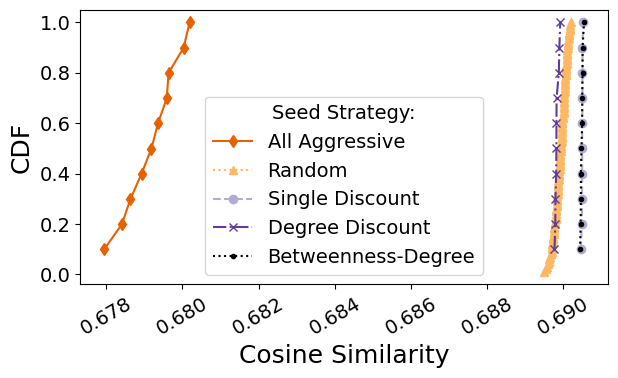}
			\caption{$WS$: Jaccard; $AC$: Top}
			\label{fig:ic_seed_strategy_J}
		\end{subfigure}
		\caption{Cosine similarities for $IC$ based on various seed strategies, Weighting Schemes ($WS$) and Activation Criteria ($AC$). Omitted combinations presented similar behavior.}
		\label{fig:ic_seed_strategy}
	\end{figure}
	
	In the next paragraphs, we present and analyze the results for selecting
	1) Seed set strategy,
	2) Weighting scheme, and
	3) Activation criterion (for $IC$) and Threshold strategy (for $LT$).
	
	\subsubsection{Seed Selection Strategy}
	
	For the rest of the experimental process, we distinguish the results of $IC$ and $LT$-based models due to the different presentation processes. Experimenting on different seed sizes (budget $k$) shows no actual accuracy implications. Instead, it only affects the duration of the diffusion process. Specifically, increasing (decreasing) the seed size just forces the process to terminate sooner (later).
	For this reason, seed size is set to $5594$, to match the number of aggressive users according to the definition presented in subsection~\ref{sec:seedselect}. Regarding seed strategy, \textit{Top Aggressive} is unnecessary as the selected seed size allows the use of all aggressive users.
	
	Figure~\ref{fig:ic_seed_strategy} presents the results regarding $IC$-based models.
	For brevity, the exhaustive list of experiments is not presented, as similar patterns were observed.
	From these results, we note a small, yet statistically significant dominance of the network-feature strategies, \textit{Single Discount} ($SD$), and \textit{Degree Discount} ($DD$). 
	Additionally, for \textit{Power}- and \textit{Weighted}-based graphs, $SD$ and $DD$ do not have significant differences.
	However, in \textit{Jaccard}, $SD$ is prevalent and also $BD$ depicts analogous performance.
	Overall though, $BD$ presents inconsistent behavior across the different weighting schemes. 
	Hence, for the $IC$ models, $SD$ constitutes the appropriate seed strategy due to its reduced computation cost compared to $DD$, and its consistently superior performance compared to $BD$. The same observation is also applicable to the $LT$-based configurations (Table~\ref{table:lt_results}). It is noted, however, that $DD$, although applicable, is not compatible with the theoretical background of the $LT$ model. Last, it is noted that $Random$ is also a viable strategy as it achieves similar, yet statistically inferior results, but leads to a computationally lightweight seed selection process.
	
	
	\subsubsection{Weighting Scheme}
	
\begin{figure}[t]
		\begin{minipage}[t]{0.75\textwidth}
			\centering
			\begin{subfigure}[t]{0.32\textwidth}
				\centering
				\includegraphics[width=\textwidth]{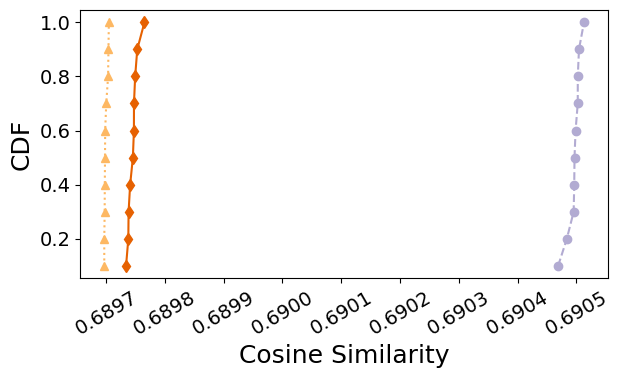}
				\caption{$AC$: Random}
				\label{fig:ic_weight_r}
			\end{subfigure}
			\hfill
			\centering
			\begin{subfigure}[t]{0.32\textwidth}
				\centering
				\includegraphics[width=\textwidth]{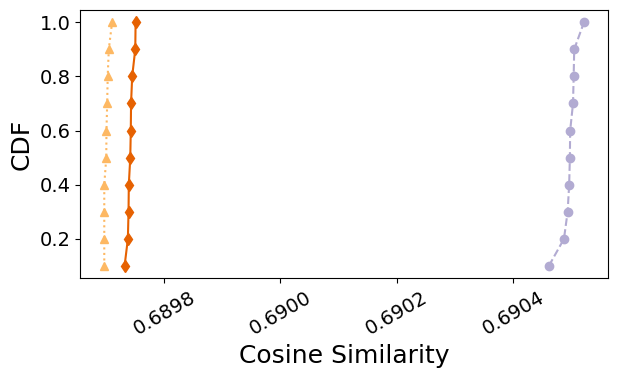}
				\caption{$AC$: Top}
				\label{fig:ic_weight_t}
			\end{subfigure}
			\hfill
			\centering
			\begin{subfigure}[t]{0.32\textwidth}
				\centering
				\includegraphics[width=\textwidth]{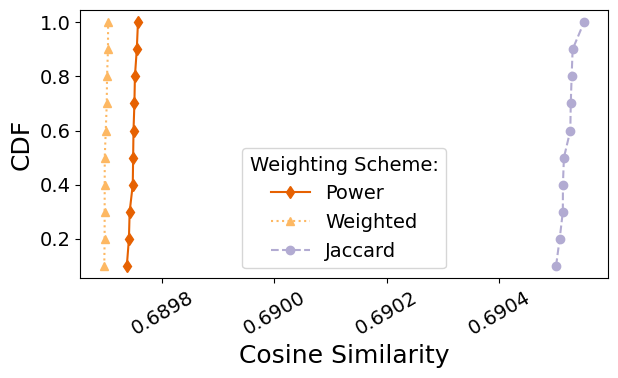}
				\caption{$AC$: Cumulative}
				\label{fig:ic_weight_c}
			\end{subfigure}
			\caption{Cosine similarities for $IC$ based on three Activation Criteria ($AC$): Random, Top and Cumulative. The selected Seed strategy is $SD$.}
			\label{fig:ic_weight}
		\end{minipage}
		\begin{minipage}[t]{0.24\textwidth}
			\centering
			\includegraphics[width=\textwidth]{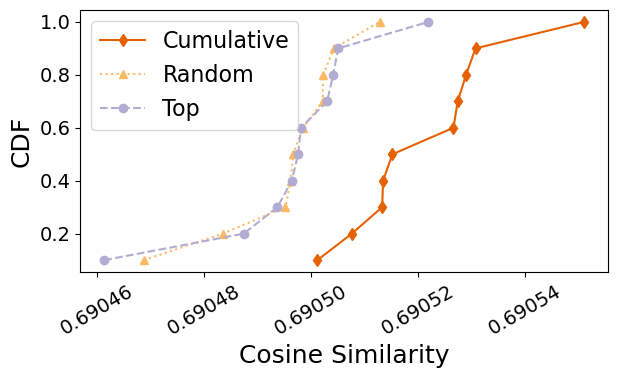}
			\caption{Cosine similarities for $IC$ based on \textit{Jaccard}. Seed strategy: $SD$.}
			\label{fig:ic_activation}
		\end{minipage}
	\end{figure}

	Given the selected seed strategy $SD$, we now analyze the results on different weighting schemes.
	Figure~\ref{fig:ic_weight} presents the results of the $IC$-based models while Table~\ref{table:lt_results} does so for the $LT$ ones. It is shown that for both $IC$ and $LT$, \textit{Jaccard} is the best weighting scheme, regardless of the activation criterion, also confirmed statistically with ANOVA and Tukey tests. This means that a user's social circle and therefore its neighborhood similarity is more appropriate to model online aggression instead of its power and credibility per se. Additionally, \textit{Weighted} and \textit{Power} as weighting schemes do not present any significant differences constituting them inappropriate weighting schemes. Thus, we conclude with \textit{Jaccard} as the appropriate weighting scheme.
	
	\subsubsection{Activation Criterion and Threshold Strategy}
	
	Up to now, we investigated the various possible configurations from a macroscopic point of view, to conclude to $SD$ as the best seed selection strategy and \textit{Jaccard} as the best weighting scheme.
	Next, we look into activation criteria and threshold strategies to select best possible setups.
	Specifically, Figure~\ref{fig:ic_activation} shows that \textit{Cumulative} activation strategy is the most dominant, while there is no clear distinction between \textit{Top} and \textit{Random} strategies, validated by ANOVA and Tukey's HSD test too.
	This prevalence is explained by the crucial effect that a user's neighborhood has on them, as it is also pointed out by the \textit{General Aggression Model}~\cite{allen2017general}.
	
	Proceeding to $LT$ models, Table~\ref{table:lt_results} shows that \textit{Aggression}-based thresholds perform slightly better and, hence, they should be used to model aggression propagation with respect to \textit{Jaccard} and $SD$. This observation is intuitive, as it suggests that the higher the aggressiveness of a user, the easier to propagate it - enabling the overall aggression diffusion process over $LT$.

	\subsubsection{Snapshot Evolution}
	
\begin{figure}[t]
		\centering
		\includegraphics[width=0.49\textwidth]{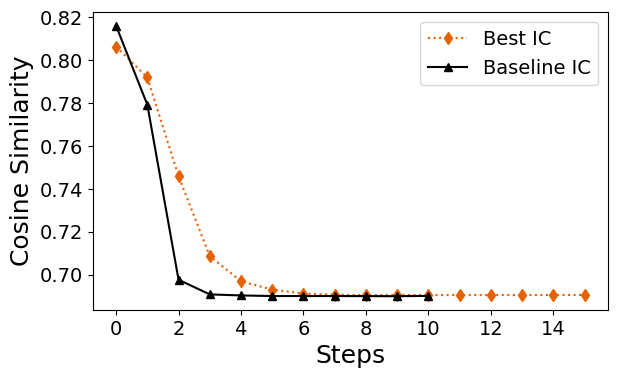}
		\centering
		\includegraphics[width=0.49\textwidth]{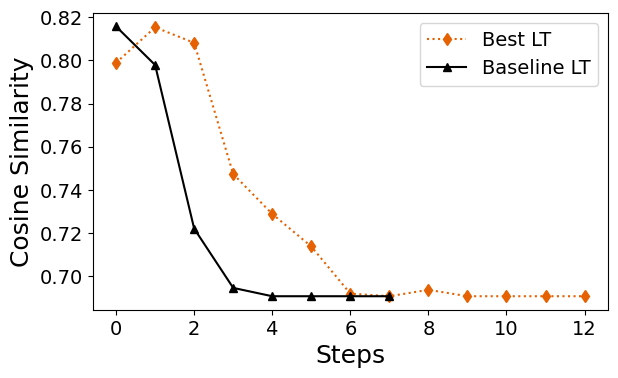}
		\caption{Snapshot evolution for the best $IC$ (left) and $LT$ (right) configurations and comparison to the random baseline models}
		\label{fig:best}
	\end{figure}

	In the previous experimental cases, the diffusion process was let to unfold until completion (i.e., after all possible time steps), and then the similarity with ground truth was calculated. However, to investigate if there is an intermediate time step (snapshot) in the propagation that can better match the ground truth change of state, we created snapshots of the diffusion process at the end of every time step and tracked the corresponding results. Additionally, in order to assess the validity and superiority of the aggression-related mechanics of the best models, a comparison to IC and LT baseline models is also conducted, presented in Figure~\ref{fig:best}. In particular, the IC baseline is characterized by a random activation probability, random seed node selection and random aggression transmission, which is needed to capture the actual performance. Analogously, the LT baseline is based on random node weights to use as thresholds and random seed node selection as well. This randomness leads to the report of their mean performance out of 10 experiment repetitions.
	
	In both $IC$ and $LT$ aggression-based models, the performance decreases rapidly within the first few steps and gradually stabilizes in the last ones. This trend shows that aggression probably does not propagate that deeply into the network (remember that steps here mean graph hops). 
	Interestingly, the stable behavior for the last steps validates the integrity of the convergence value studied earlier. Experimenting with two more similarity metrics (Pearson and Spearman correlations), it was found that Pearson behaves like Cosine, but Spearman presents an early increase with a steady drop afterwards. However, Cosine was preferred for our analysis, since the normality requirement for Pearson is not satisfied and Cosine is more sensitive to subtle changes than Spearman.
	
	Moreover, the comparison to the baselines shows that - for both IC and LT models - adopting aggression-related mechanics in the weighting scheme, seed selection process and activation/threshold strategy offers additional value and enhances the performance of the modeling. This means that the original diffusion models, although applicable, do not have enough expressive power to capture and describe the complex nature of online aggression spreading. It also suggests that the proposed mechanics are promising and towards the right direction to better understand aggression diffusion over Twitter and social networks in general.
	
	\textbf{Takeaways:} Summarizing, aggression diffusion over Twitter can be described by a cascading model, such as $IC$ or $LT$, fulfilling the first part of \textit{Contributions~\ref{contribution2}} and\textit{~\ref{contribution3}}. Considering that the original $IC$ and $LT$ are not able to capture online aggression dynamics, new mechanisms have to be created. For both models, the best seed strategy is $SD$, also supported by statistical significance, meaning that the diffusion process should initiate from the most central nodes in the network, e.g. users with many friends and followers around whom large user communities are formed. However, for $LT$ one has the luxury of sacrificing the model's stability to gain in terms of computational efficiency, by selecting the trailing but still decently performing $Random$ seed strategy.
	Moreover, for both cases, the best performing weighting scheme is \textit{Jaccard} which expresses that relations are formed based on users' neighborhood similarity. The explanation of this outcome is that to predict aggression diffusion one should examine the social interactions instead of individual user attributes. Specific to $IC$-based models, activation criterion should be set to \textit{Cumulative}, enabling the whole neighborhood of a node to affect its aggression state, while for $LT$-based models, an \textit{Aggression}-based threshold strategy is preferred, as it enables the overall dissemination process. Lastly, the diffusion process should unfold until convergence to achieve a stable state.
	
	\subsection{Case Study: modeling aggression diffusion}\label{sec:case-study}
	
	To evaluate the best $IC$ and $LT$ models (as concluded earlier), we apply them to the SL Twitter dataset. Here, we remind the reader that SL contains \textit{normal} and \textit{aggressive} labels for two network snapshots: 2016 and 2019. Using the user labels of the first snapshot, we run the $IC$ and $LT$ models and measure the $AUC$ of the predicted labels, vs. the ones of the second snapshot.
	\begin{table}[t]
		\centering
		\caption{Proposed $IC$ and $LT$ models compared with $OD$ models~\cite{terizi2020angry} on the SL dataset.}
		\label{table:od_comparison}
		\begin{tabular}{c | c | c | c | c }
			\hline
			\textbf{Model} & Deffuant\_P & HK\_1.0\_P & IC & LT \\
			\hline \hline
			\textbf{AUC} & 0.79 & 0.76 & 0.82 & \textbf{0.89}\\
		\end{tabular}
	\end{table}
	The $AUC$ achieved by $IC$ (with $SD$, Jaccard, and Cumulative) is $0.82$, while for $LT$ (with $SD$, Jaccard and Aggression) is $0.89$. Thus, both configurations perform well in modeling aggression diffusion, and $LT$ seems to provide slightly better performance, with fewer false positives (aggressors). 
	
	To support the eligibility of $IC$ and $LT$ as diffusion processes that can capture online aggression diffusion, Table~\ref{table:od_comparison} presents the results of our $IC$ and $LT$ models in contrast to the OD models introduced in~\cite{terizi2020angry}. Specifically, the authors in~\cite{terizi2020angry} report only their top-performance models, i.e.,~that achieve more than $0.70$ in $AUC$, namely $Deffuant$ and $HK$. As it is shown, both $IC$ and $LT$ best configurations outperform the OD alternatives.
	
	\subsection{Aggression Minimization}\label{sec:aggression-minimization-experiments}
	
	The second part of the experimental process pertains to aggression minimization ($CAM$ and $BAM$ problems) and aids in \textit{Contributions~\ref{contribution2}} and\textit{~\ref{contribution3}}. To address the $CAM$ problem, a competitive cascade process is exploited aiming at decreasing the final aggression score of the network, while for $BAM$, the proposed node (edge) blocking mechanisms select the nodes (edges) to remove before the launch of the aggressive cascade.
	
	\subsubsection{Competitive Aggression Minimization}
	
	With respect to $CAM$, the positive cascade should be similar to the negative one to compete in equal terms, i.e., for both $IC$ and $LT$ models, and both positive and negative cascades the chosen weighting scheme is \textit{Jaccard}. For the $IC$, the activation strategy is \textit{Cumulative}, and for $LT$ the threshold strategy is \textit{Aggression}. The experimentation comprises of the different seed strategies for the positive cascade (\textit{All Aggressive}, \textit{Top Aggressive}, $SD$, $DD$ and $BD$), as well as the various healing mechanisms (\textit{Vaccination}, \textit{Aggression transfer}, \textit{Decaying transfer} and \textit{Hybrid}).
	
\begin{figure}[t]
		\centering
		\begin{subfigure}[t]{0.32\textwidth}
			\centering
			\includegraphics[width=\textwidth]{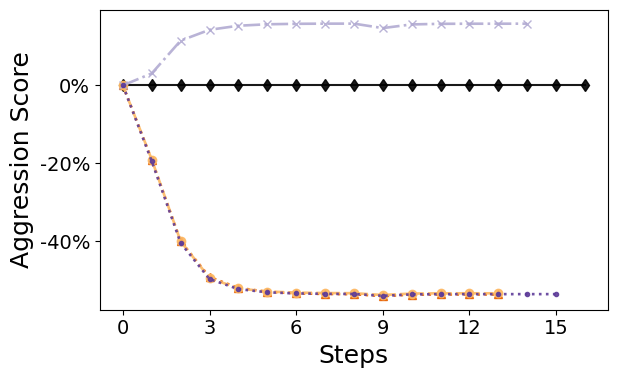}
			\caption{\textit{All Aggressive} seed strategy}
			\label{fig:ic_agg_evo_min_aa}
		\end{subfigure}
		\hfill
		\centering
		\begin{subfigure}[t]{0.32\textwidth}
			\centering
			\includegraphics[width=\textwidth]{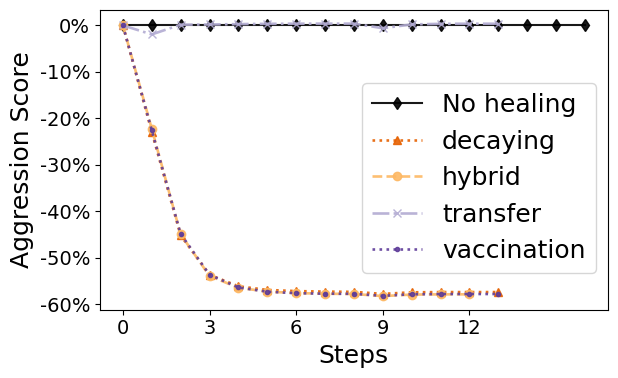}
			\caption{\textit{Random} seed strategy}
			\label{fig:ic_agg_evo_min_r}
		\end{subfigure}
		\hfill
		\centering
		\begin{subfigure}[t]{0.32\textwidth}
			\centering
			\includegraphics[width=\textwidth]{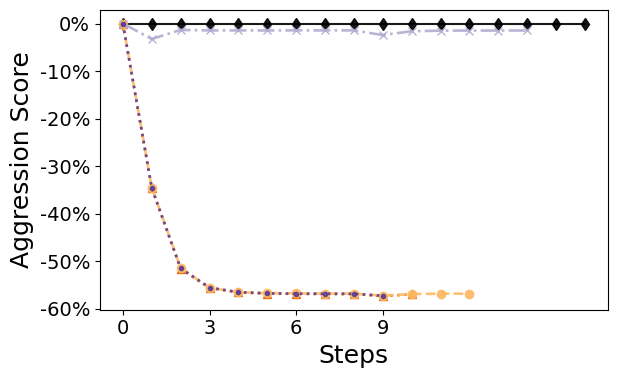}
			\caption{$DD$ seed strategy}
			\label{fig:ic_agg_evo_min_dd}
		\end{subfigure}
		\medskip
		\begin{subfigure}[b]{0.32\textwidth}
			\centering
			\includegraphics[width=\textwidth]{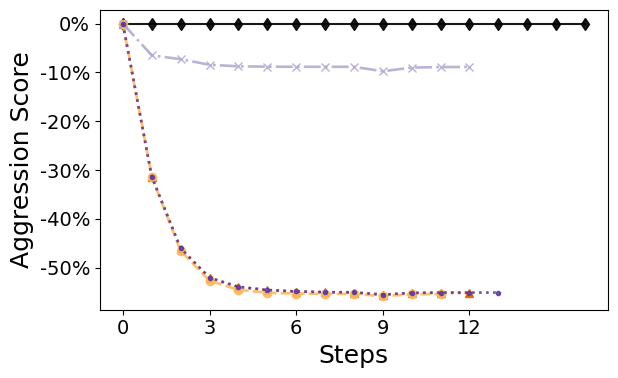}
			\caption{$SD$ seed strategy}
			\label{fig:ic_agg_evo_min_sd}
		\end{subfigure}
		\quad
		\begin{subfigure}[b]{0.32\textwidth}
			\centering
			\includegraphics[width=\textwidth]{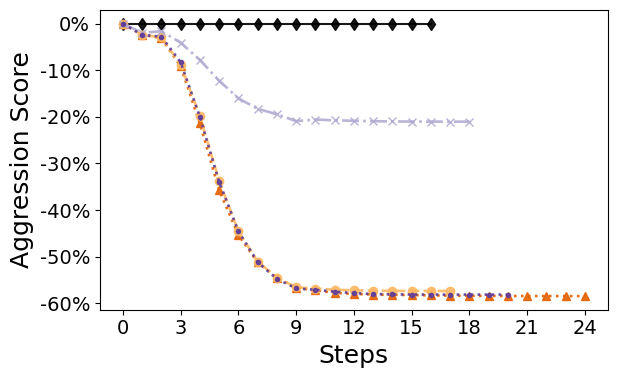}
			\caption{$BD$ seed strategy}
			\label{fig:ic_agg_evo_min_bd}
		\end{subfigure}
		
		\caption{Healing strategy comparison on $IC$ models, under $CAM$ problem.}
		\label{fig:ic_agg_evo_min}
	\end{figure}

	Here, we focus on the best configurations of the modeling phase mentioned earlier.
	Figure~\ref{fig:ic_agg_evo_min} presents aggression evolution during the competitive cascade process, for various seed and healing strategies on the $IC$ models. Y-axis presents the ratio of loss or gain with respect to the case where no healing is applied. First, regarding the seed strategy of the positive cascade, $BD$ presents the best results for every healing mechanism reaching 61.2\% aggression reduction, while \textit{Random} and $DD$ follow with $\sim$$57\%$, whereas $SD$ and \textit{All Aggressive} come last. Thus, $BD$ is the best performing and also stable option, while \textit{Random} is a viable alternative when computational cost is important.
	
	With respect to the healing strategies, regardless of the positive seed strategy, \textit{Transfer} presents the worst performance with a best-case reduction of $\sim$$20\%$, due to the simple aggression score transferring, no matter the result on the corresponding node. The rest of the healing mechanisms perform similarly, irrespective of the seed strategy. In particular, \textit{Vaccination} is the most dominant mechanism in all seed strategies achieving 61.2\% aggression reduction, apart from $BD$ where \textit{Decaying transfer} is equally efficient. For the majority of the seed strategies though \textit{Decaying transfer} and \textit{Hybrid} follow \textit{Vaccination} with $\sim$57\%. These results are intuitive, since \textit{Vaccination} makes the strong assumption that nodes get completely healed, whereas \textit{Decaying transfer} is a more relaxed and realistic version of \textit{Vaccination}, and \textit{Hybrid} lies in between them. Last, the odd case of $BD$ - where \textit{Decaying transfer} and \textit{Vaccination} are equally strong - implies that the specific seed strategy for the positive cascade is so efficient, that the actual differences in the healing mechanisms are of lesser importance.
	
	Another critical factor of a healing strategy's efficiency is the number of nodes that get activated by the negative cascade, since the same aggression reduction on a larger number of activated nodes expresses a more efficient mechanism. In particular, all healing mechanisms, except \textit{Transfer}, activate about the same number of nodes (plots omitted for brevity). Combining the results on aggression reduction with the number of activated nodes, we conclude that \textit{Vaccination} is the best healing mechanism regardless of the seed strategy, but due to its mostly theoretical nature, \textit{Decaying transfer}, in combination with $BD$, seem to be the most efficient strategies for aggression minimization in $IC$-based models.
	
\begin{figure}[t]
		\centering
		\begin{subfigure}[t]{0.49\textwidth}
			\centering
			\includegraphics[width=\textwidth]{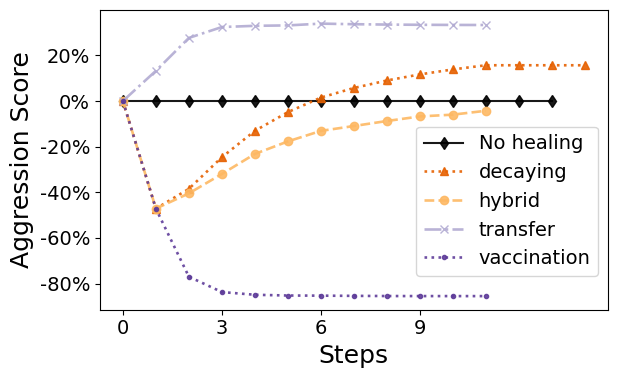}
			\caption{\textit{All Aggressive} seed strategy}
			\label{fig:lt_agg_evo_min_aa}
		\end{subfigure}
		\hfill
		\centering
		\begin{subfigure}[t]{0.49\textwidth}
			\centering
			\includegraphics[width=\textwidth]{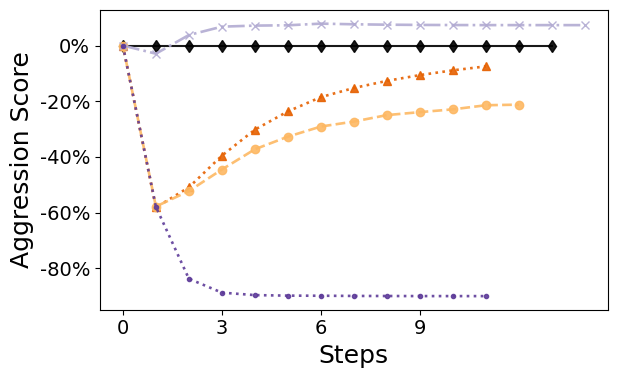}
			\caption{\textit{Random} seed strategy}
			\label{fig:lt_agg_evo_min_r}
		\end{subfigure}
		\medskip
		\begin{subfigure}[b]{0.49\textwidth}
			\centering
			\includegraphics[width=\textwidth]{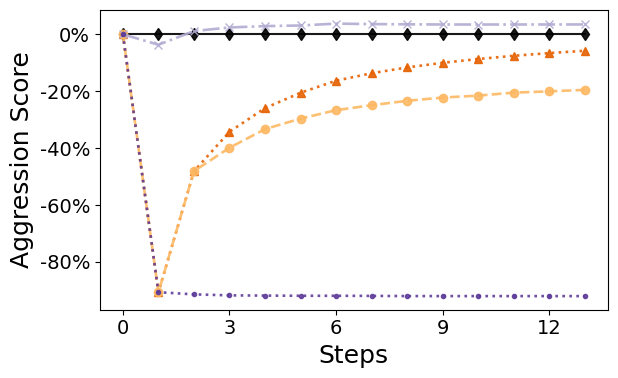}
			\caption{$SD$ seed strategy}
			\label{fig:lt_agg_evo_min_sd}
		\end{subfigure}
		\hfill
		\centering
		\begin{subfigure}[b]{0.49\textwidth}
			\centering
			\includegraphics[width=\textwidth]{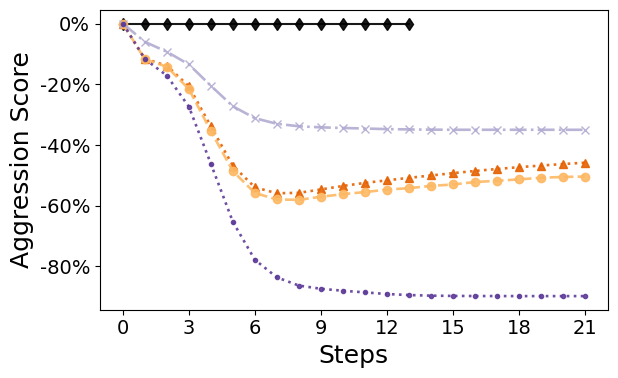}
			\caption{$BD$ seed strategy}
			\label{fig:lt_agg_evo_min_bd}
		\end{subfigure}
		
		\caption{Healing strategy comparison on $LT$ models, under $CAM$ problem}
		\label{fig:lt_agg_evo_min}
	\end{figure}

	Proceeding to the $LT$ models, it is mentioned that $DD$ is not a viable option as it has been explained already. With this in mind, Figure~\ref{fig:lt_agg_evo_min} presents that $BD$ is dominant, achieving up to 92\% reduction, while $SD$ and \textit{Random} closely follow with 91\% and 90\%, respectively. Moving to the healing strategies, \textit{Vaccination} achieves the best reduction of approximately 92\%, but, in contrast to the $IC$ case, it is the only one that preserves the reduction during the whole process, as it completely deactivates the affected nodes, and, thus, they have zero effect on their neighbors during the subsequent steps. On the contrary, the rest of the healing strategies can not preserve the reduced aggression levels, suggesting a potential inefficiency of competitive cascades on $LT$ models, with the exception of $BD$ where the employed mechanisms are able to retain their aggression reduction levels.
	Hence, it is observed that in $LT$ models, $BD$ is of vital importance to the effect of the minimization process, while a realistic reduction of $\sim$$50\%$ is possible when \textit{Hybrid} or \textit{Decaying transfer} are used as the healing strategies.
	
	Regarding the activated nodes during each propagation process, \textit{Vaccination} activates the least number of nodes, while \textit{Decaying transfer} stimulates 2x more nodes compared to \textit{Hybrid} in most cases. Also, considering the theoretical nature of \textit{Vaccination}, \textit{Decaying Transfer} should be prioritized as a healing mechanism.
	
	\subsubsection{Blocking Aggression Minimization}
	
	Moving to the second minimization problem ($BAM$), the results of four methods combining: 1) node or 2) edge blocking (removal), while using a) \textit{Jaccard} as edge weight in the adjacency matrix \textbf{A}, or b) the modified version with the aggression scores (please see details in Sec.~\ref{sec:aggression-minimization-experiments}) are briefly discussed and compared to a no-immunization process as baseline.
	For node and edge blocking, NetShield~\cite{tong2010vulnerability} and NetMelt~\cite{tong2012gelling} algorithms were used, respectively, to select nodes or edges to remove.
	
	For $IC$, the aggression-based variations do not present significant improvement over the baseline.
	Aggression reduction is possible - although smaller than competitive cascades - by exploiting node removal, achieving a reduction of 8\%. We note that the number of removed nodes is $5594$, equal to the seed size of the competitive cascade process. However, when edge blocking is examined, the overall aggression score is, in fact, doubled, regardless of the type of matrix \textbf{A}, due to the combination of significant edge removal and the degree-based seed strategy of the underlying diffusion process. Moreover, to let \textit{NetMelt} reach the same number of nodes as \textit{NetShield}, $\sim$$470k$ edges have to be removed, which is unrealistic.
	
	For $LT$, edge removal (with both types of \textbf{A}) presents a switching behavior of 1\% increase and decrease in overall aggression, in relation to the baseline of no immunization, leading to an insignificant outcome on aggression minimization. Node removal, though, performs better, since the reduction reaches 6\% for the \textit{Jaccard}-based \textbf{A}, and 8\% for the aggression-based case. In general, the aggression alternative of the adjacency matrix seems to benefit the process, but more in-depth research is needed to understand its trade-offs. Lastly, regarding the number of activated nodes (results omitted), in comparison to the competitive cascades, the immunization methods here tend to activate 4x times more nodes overall.
	
	\textbf{Takeaways:} Aggression minimization in Twitter through competitive cascades outperforms blocking techniques. Specifically, using $IC$ models a reduction of 61.2\% with respect to the total initial aggression score of the network is achieved, in contrast to the 8\% of the blocking methods. Similarly, $LT$ models are able to reach an aggression reduction of 50\%.
	The employment of $BD$ seed strategy is of crucial importance to the overall minimization process. Particularly, it is the reason behind the high aggression reduction in both $IC$ and $LT$ models, while specifically for $LT$ it constitutes the only option for the seed strategies as the rest lead to instabilities of the overall reduction. Last, it should be noted that even if blocking mechanisms were undisputedly superior to competitive cascades in terms of aggression reduction, they would be still too invasive. This means that they disrupt the network heavily by completely removing nodes and edges, as it is also observed in Twitter~\cite{milosavljevic2016banning}. Therefore, even with the proper selection of users to block, the effects of this action are observable by many users and disrupt their experience. These observations cover the second and last part of \textit{Contributions~\ref{contribution2}} and\textit{~\ref{contribution3}}.
	
	\section{Discussion \& Conclusion}\label{sec:future}
	
	In this work, the dual problem of aggression modeling and aggression minimization in Twitter was examined. To the best of our knowledge, aggression modeling has been addressed only in~\cite{terizi2020angry}, where it was approached through opinion dynamics instead of cascading models. Both approaches present comparable performance in modeling real Twitter data on aggression diffusion, while having some distinguishing features regarding the process and convergence conditions.
	We believe our work advances the community's understanding on aggression diffusion one step further, while determining the most prevalent of the two methods is left for a follow-up work. Apart from aggression modeling, however, we are the first to address aggression minimization. We exploit models from the domains of competitive cascades and immunization and discover strategies to implement them as minimization methods.
	
	\noindent
	\textbf{Results:}
	First, we showed that $IC$ and $LT$ are suitable for modeling aggression diffusion, as they can reach a cosine similarity of $\sim$70\% with ground truth data. 
	The results showed that by focusing on the most central users - instead of the most aggressive ones - to initiate the process, online aggression diffusion in Twitter can be better modeled. Additionally, capturing neighborhood similarity is vital for describing aggression dissemination, hence \textit{Jaccard} overlap should be used as the weighting scheme for graph edges. Moreover, \textit{Cumulative} activation criterion of $IC$ engulfs the team influence on information propagation, while for $LT$, \textit{Aggression} threshold strategy suggests that high aggressiveness propagates easier, enabling aggression diffusion.
	
	With respect to aggression minimization, competitive cascades achieved a total of 61.2\% reduction in $IC$, and 50\% in $LT$ models. Specifically for $IC$, the similar behavior of the most prevalent seed strategies allows a trade-off between stability ($BD$) and computational cost (\textit{Random}), while for $LT$, $BD$ is a one-way option as the rest of the strategies are unstable. Moreover, \textit{Decaying Transfer} was the most appropriate healing mechanism, despite \textit{Vaccination}'s general higher performance. This is because \textit{Vaccination} naively presumes that positively affected users, e.g., through a cyberbullying sensitizing campaign, will not switch back to abusive behavior. Node blocking methods reached $\sim$8\% reduction in both $IC$ and $LT$. Edge blocking is somewhat inappropriate, as it interferes with the degree of nodes making it incompatible with the degree-based $BD$ seed strategy of the aggression cascade. In any case, competitive cascades with $BD$ and \textit{Decaying transfer} as seed and healing strategy respectively, outperform the blocking mechanisms.
	
	\noindent
	\textbf{Implications of our work:}
	Next, we unfold and discuss what the real-world implications of our work are. Considering the case of a Twitter community moderator, the above observations provide some tools to control and provision the community's function. That is, to ensure the interception of aggression spread, such moderators should closely monitor the online behavior of central and influential users. This is in case they turn aggressive, as they could initiate a burst of aggression increase in their community and Twitter as a whole. However, if aggression is already established in a specific community, the moderators should regulate the activity of users sharing many common friends and followers, with the already aggressive ones. Lastly, to measure and avoid underestimating the impact of aggressive activity, they should adopt a holistic approach by considering the cumulative effect of one's neighbors on their aggressive behavior.  
	
	Regarding the observations on competitive cascades and blocking techniques, and to reduce aggression in Twitter, we discuss what would the actual real world recommendations be. First, it is clear by now that the already used methods of banning users - and by extension removing multiple edges - are too invasive and disrupt the network and user experience. This fact may even be the reason behind Twitter's reluctance to ban users, even when they are evidently aggressive and disruptive. Instead, our results show that a simple awareness campaign (e.g., sharing posts for eliminating racism or hate speech) can go a long way, as it could both be less intrusive and lead to larger aggression reduction.
	Even better, community moderators could focus such campaigns on Twitter users with a crucial role in the flow of information that connect distinct parts of the network (community ambassadors) to reduce the time needed to reach users who are in the periphery of the communities, as in this way \textit{Decaying Transfer} (the best healing strategy), would operate on its maximum effect. Last but not least, these campaigns could help users improve their behavior in other social media as well, an outcome that is not ensured by blocking mechanisms.
	
	\noindent
	\textbf{Limitations and future work:}The methods presented in this work consist the starting point of understanding and addressing online aggression diffusion and minimization in social networks in general, and in Twitter, specifically.
	Therefore, instead of providing results on state-of-the-art methods, the aim here was to present a proof of concept for online aggression modeling and minimization using basic $IC$ and $LT$ methods, focused on Twitter.
	As this is a multi-faceted and dynamic phenomenon, the results of this work should be extended to other social media platforms (apart from Twitter), and investigated in greater depth to develop more straightforward and realistic online aggression counter measures.
	To this end, we steer our focus for future work towards the below promising research directions:
	
	\begin{enumerate}
		\item Explore further the possible weighting schemes, seed strategies, activation/threshold criteria and aggression transferring mechanisms, to better capture aggression dynamics on real networks and different datasets, and especially in aggression-infused networks such as Gab~\cite{zannettou2018gab, mathew2019spread} and Voat\footnote{https://www.voat.xyz/}.
		\item Investigate network-ubiquitous factors such as user anonymity and freedom of speech that enable or complicate an epidemic spread of aggression under different virus-spreading models to better understand its real-world implications.
		\item Examine ways to incorporate offline blocking methods into the diffusion process and dynamically minimize aggression, while monitoring the obtrusiveness of these methods, as they seem appropriate for LT-based aggression diffusion.
	\end{enumerate}
	
	Many challenges lurk in addressing the above directions. However, all of these future steps are feasible and promote our knowledge on addressing online aggression once and for all.
	
	\section*{Acknowledgements}
	
	This research has been partially financed by the European Union and Greek national funds through the Operational Program	Competitiveness, Entrepreneurship, and Innovation, under the call RESEARCH - CREATE - INNOVATE (Project Code: T2EDK-03898), as well as from the H2020 Research and Innovation Programme under Grant Agreement No. 875329 (LifeChamps), No. 830927 (Concordia), and the Marie Skodowska-Curie under Grant Agreement No. 691025 (ENCASE).
	The presented work reflects the authors views solely.

	\bibliographystyle{ACM-Reference-Format}
	\bibliography{main}
	
\end{document}